# All-optical mode unscrambling on a silicon photonic chip

Andrea Annoni[1], Emanuele Guglielmi[1], Marco Carminati[1], Giorgio Ferrari[1],
Marco Sampietro,[1] David A. B. Miller[2], Andrea Melloni[1], Francesco Morichetti[1]

Corresponding author: francesco.morichetti@polimi.it

[1]*Dipartimento di Elettronica, Informazione e Bioingegneria, Politecnico di Milano,*
*via Ponzio 34/5, 20133 Milano, Italy*

[2]*Ginzton Laboratory, Stanford University, Spilker Building, 348 Via Pueblo Mall, Stanford CA 94305, USA*

**Abstract**

Propagation of light beams through scattering or multimode systems may lead to randomization of the spatial coherence of the light. Although information is not lost, its recovery requires a coherent interferometric reconstruction of the original signals, which have been scrambled into the modes of the scattering system. Here, we show that we can automatically unscramble four optical beams that have been arbitrarily mixed in a multimode waveguide, undoing the scattering and mixing between the spatial modes through a mesh of silicon photonics Mach-Zehnder interferometers. Using embedded transparent detectors and a progressive tuning algorithm, the mesh self-configures automatically and reset itself after significantly perturbing the mixing, without turning off the beams. We demonstrate the recovery of four separate 10 Gbits/s information channels, with residual cross-talk between beams of -20dB. This principle of self-configuring and self-resetting in optical systems should be applicable in a wide range of optical applications.

When a coherent light beam passes through an optical object, interference from scattering or different paths can distort the beam. Strong diffuse scatterers[1,2] and even simple multimode fibres or waveguides[3,4] can generate complex speckle patterns from simple beams, giving strong scrambling of multiple beams and any information on them[5]. For beams of the same wavelength and polarization, historically we have had no good way to separate these beams and channels optically. This problem is worse if we do not know the characteristics of the optical object, and worse still if the object changes in time.

If we measure the object interferometrically or use some global feedback algorithm we can calculate the input field required for a desired output field[1,2,4]. A spatial light modulator can set up any one such input field from an input beam, but it cannot simultaneously construct multiple arbitrary overlapping input fields from multiple inputs. In few-mode optical fibres or waveguides, specific modes can be separated, based on their symmetries and/or different phase velocities[6-10], and signals on well-defined modes can be interchanged or switched[11,12]. But, if we use arbitrary orthogonal input beams and/or the beams couple or scatter during propagation because of imperfections or bends, such approaches cannot generally separate the resulting complex superpositions of output guided modes and information. Coherent detection, together with analog-to-digital conversion and digital electronic MIMO processing, can deconvolve the separate channels again[13,14], up to electronic bandwidths, and can compensate different group delays of different modes[13]. But, these approaches require complex digital circuits with associated power, speed and capacity limits.

In waveguides where the loss is essentially equal on all the different propagating modes, in principle scattering between modes can be undone with a unitary linear processor (or, in practice, one with uniform loss across all channels). A triangular mesh of 2x2 tuneable beam splitters (Fig. 1a) is a well-known architecture implementing arbitrary unitary operations[15] (See Supplementary Note 1 for discussion of other meshes). Such meshes were successfully implemented in integrated photonics for quantum applications[16-18], but progressive self-configuring algorithms[19-22] were not yet available, leading to time-consuming global calibration and optimisation algorithms. Self-configuration of a triangular mesh recently demonstrated automatic coupling of one arbitrary input beam and automatic rerouting of a single signal through an optical switching matrix[23].

Here, we demonstrate that strong mixing between modes can be undone all-optically, automatically and without any advance knowledge of the mixing object's details; furthermore, our approach can adapt in real time to changes in the object. Explicitly, we show the separation of four optical beams after they have been completely and arbitrarily mixed in a multimode waveguide. We use a silicon photonic mesh architecture[15] with built-in transparent contactless integrated photonic probe (CLIPP) detectors[24] and progressive self-configuring algorithms[19-22] to demonstrate simultaneous unscrambling, sorting and tracking of the four beams, and information recovery from four mixed mode-division-multiplexed (MDM) channels. We also then strongly perturb the guide so the channels are completely mixed again, and show automatic recovery of the separation in seconds.

# Results

**Self-configuration of the mesh**

An $N{\times}N$ triangular array or mesh of $N(N\text{-}1)/2$ tuneable $2{\times}2$ beam splitters ($S_{m,k}$) is connected according to the mesh and photodetector topology proposed in[19,20] (Fig. 1a). This mesh enables the factorization of an arbitrary $N{\times}N$ unitary transformation, described by a transmission matrix $\mathbf{H}_{mesh}$, into a sequence of simple $2{\times}2$ unitary transformations[15,20]. The evolution of the optical field $\mathbf{E}$ along the mesh,

$$\mathbf{E}_{m,k+1} = \mathbf{T}_{m,k}\mathbf{E}_{m,k} = \mathbf{T}_{m,k}\mathbf{T}_{m,k-1}\mathbf{E}_{m,k-1} = \prod_{q=0}^{k-1}\mathbf{T}_{m,k-q}\mathbf{E}_{m,1}, \tag{1}$$

can be described[15] by the product of $N{\times}N$ transmission matrices $\mathbf{T}_{m,k}$ associated with each beam splitter $S_{m,k}$, where $m = 1,2, \ldots N\text{-}1$ is the progressive index of the mode reconstructed at the output port $Out_m$ and $k = 1,2, \ldots N\text{-}m$ is the $k$-th step performed to extract the $m$-th mode (see also Ref. [20] for an equivalent statement of this mathematics). Each matrix $\mathbf{T}_{m,k}$ is an $N{\times}N$ matrix ($4{\times}4$ in Fig. 1a) constructed by starting with an identity matrix and then replacing the elements $T_{ij}$ ($i,j= N\text{-}k, N\text{-}k+1$) with the elements of the $2{\times}2$ matrix $\mathbf{T}_{BS}$ of the tuneable beam splitter $S_{m,k}$, given[15], for example, by

$$\mathbf{T}_{BS} = -je^{j\frac{\phi_2}{2}}\begin{pmatrix} -\sin\frac{\phi_2}{2} & e^{j\phi_1}\cos\frac{\phi_2}{2} \\ \cos\frac{\phi_2}{2} & e^{j\phi_1}\sin\frac{\phi_2}{2} \end{pmatrix}, \tag{2}$$

where $\phi_1$ and $\phi_2$ are controllable phase shifts [in Eq. (2) subscripts $m,k$ are omitted for notational simplicity]. Each beam splitter $S_{m,k}$ modifies only the components $N\text{-}k$ and $N\text{-}k+1$ of the field vector $\mathbf{E}_{m,k}$, leaving the other $N\text{-}2$ components unchanged. Importantly, at each stage of the mesh a single transparent light detector (placed on either the waveguide $N\text{-}k$ or $N\text{-}k+1$) enables us to follow the mode evolution throughout the entire mesh[20], without impairing the operation of the mesh itself.

To illustrate the self-configuration procedure, consider a first beam shining on the mesh inputs, and hence generating a vector $\mathbf{E}_{11}$ of coherent input beam amplitudes at the input ports. To output all of this beam power at port $Out_1$, the beam splitters $S_{11}$, $S_{12}$ and $S_{13}$ are progressively adjusted to null out the power at the embedded detectors[19-21]. Independent of the relative amplitudes and phases in the input ports, all the power from the input vector $\mathbf{E}_{11}$ is automatically combined into one output beam. We can consider this mathematically[20] to be a progressive multiplication by matrices $\mathbf{T}_{11}$, $\mathbf{T}_{12}$, and $\mathbf{T}_{13}$, generating vectors $\mathbf{E}_{12}$, $\mathbf{E}_{13}$, and $\mathbf{E}_{14}$, constructing an overall matrix $\mathbf{M}_1$, and progressively combining these amplitudes into the first element of $\mathbf{E}_{14}$. As unitary operators preserve orthogonality, if we now shine a second beam with an orthogonal input vector of amplitudes into the mesh, none of that beam appears in port $Out_1$. Hence all of that second beam will pass through the transparent photodetectors into beamsplitters $S_{21}$ and $S_{23}$, giving the vector of amplitudes $\mathbf{E}_{21}$ in the lower three guides; those beamsplitters can then be similarly automatically aligned to couple that second beam all to port $Out_2$, implementing an additional matrix $\mathbf{M}_2$. Configuring each $m$-th diagonal row of the mesh, which is associated with the mode reconstruction matrix

$$\mathbf{M}_m = \prod_{k=0}^{N-m-1} \mathbf{T}_{m,N-m-k} ,\qquad(3),$$

we can separate any arbitrary set of four orthogonal input vectors to the four output ports $Out_1$, $Out_2$, $Out_3$, and $Out_4$, formally implementing an arbitrary unitary transform[20]. Note such training does not require any calibration of the phase shifters inside the mesh, and can be completed automatically and progressively in one algorithmic pass through the set of beamsplitters.

Here we described turning on the training beams one by one, specifically not turning on the second beam until the row of beamsplitters $S_{11}$, $S_{12}$, and $S_{13}$ is fully configured, and similarly for further beams and beamsplitter rows. Indeed, such a separated training may always be required when working with simple continuous DC beams and for related progressive algorithms that can run based only on detectors at the output ports[19,22]. If, however, the beams of interest are not mutually coherent, and if we can put some identifying "key" on each such training beam, such as a small modulation at a different frequency or "tone" for each beam, then the whole configuration process can be run simultaneously with all the beams on at once[19].

**Photonic integrated mode unscrambler**

To demonstrate on-chip mode unscrambling, we realized a silicon photonic 4×4 triangular array of six MZIs (Fig. 1b, Fig. 1c) on a 220nm silicon-on-insulator (SOI) platform (see Methods). Phase shifts $\phi_1$ and $\phi_2$ are thermally controlled using microheaters, and CLIPP[24] detectors monitor the power at the lower output port of each MZI (see Supplementary Note 2 and Supplementary Figure 1). In the CLIPP[24], surface-state absorption (SSA)[25-27] in the silicon guide gives photoconduction that can be detected capacitively, giving a sub-bandgap all-silicon photodetector for high-sensitivity measurement of light power in the waveguide without introducing any appreciable loss, reflection or phase perturbation in the optical field[24]; the absence of any additional absorption means the CLIPP monitors the power without introducing any different absorption to disturb the mode orthogonality. For simultaneous read-out of all the CLIPPs, a custom-designed multi-channel CMOS ASIC was bridged to the silicon photonic chip and mounted on the same printed circuit board (PCB) (Fig. 1d)[28,29]. Automatic tuning and stabilization of each MZI was performed using a customized field programmable gate array (FPGA) platform connected to the photonic PCB (see Methods).

Mode scrambling is intentionally induced with an integrated mode mixer consisting of a straight multimode waveguide section with four single mode input/output waveguides (Supplementary Note 3). Four spatially decoupled input beams (modes A, B, C and D) are injected into the silicon chip through an array of four single mode fibers, which are vertically coupled to the silicon waveguides through a four-channel glass transposer (Fig. 1d). This mode mixer can be represented by a matrix **H** that maps any vector of each input in modes A, B, C, or D in Fig. 2a to some resulting vector of field amplitudes at the single-mode waveguide outputs that form the inputs to the mesh.

We separately checked an identical stand-alone mode-mixing waveguide, showing that at 1525 nm wavelength each input mode is scrambled at the output ports with about 25% power distribution (Supplementary Figure 2). The integrated mixer provides almost constant mode power coupling across a wavelength range of several tens of nanometres, with negligible differential group delay (DGD) among the different modes, and therefore emulates the mode scrambling occurring in short links of a few-mode fibre (FMF)[30].

**Sorting out mixed modes**

To illustrate the reconstruction of modes scrambled by propagation through the mode mixer, in the example of Fig. 2a, the first row of the mesh ($\mathbf{M}_1$) is progressively configured to have the optical mode D reconstructed at $Out_1$. In this experiment, all four input modes, which share the same optical wavelength $\lambda_0 = 1525$ nm, are switched on and are injected into the silicon chip with the same power of 0 dBm. Each one of the input modes A, B, C, and D was "keyed" by using external intensity modulators, adding a weak (5% peak-to-peak amplitude) modulation at frequencies $f_q = \{4$ kHz, 7 kHz, 10 kHz, 11 kHz$\}$ for each of the input modes ($q = $ A, B, C, D), respectively. Individual monitoring of the $q$-th mode was then performed by demodulating the CLIPP signals at the corresponding frequency $f_q$ (see Supplementary Note 4 and Supplementary Fig. 3)[31,32].

First, MZI $S_{11}$ (see Fig. 2b) is tuned to cancel out the power associated with mode D at the lower output port, where CLIPP1 is integrated. The map of Fig. 2c shows the intensity of mode D versus the phases ($\phi_1$, $\phi_2$) (see Fig. 2b) of $S_{11}$ as measured directly by CLIPP1. The thermal phase shifters are initially set to a non-zero value, such as $S^i_{11}(\pi,\pi)$, so as to be able to either increase or decrease the phase shift during the tuning operation. Once convergence to a local minimum $S^f_{11}$ is achieved, the procedure is sequentially repeated through the following stages $S_{12}$ and $S_{13}$. After the tuning of each beam splitter in the first row of the mesh ($\mathbf{M}_1$), the powers of the sorted mode D and of the concurrent modes A, B and C were measured at port $Out_1$ over a wavelength range of 20 nm around $\lambda_0$. Note that, although this mesh configuration process leads to a progressive increase in the output power of the reconstructed mode at $Out_1$ (Fig. 2d$_4$), the crosstalk associated with each concurrent mode does not decrease monotonically as this configuration progresses (Fig. 2d$_1$-d$_3$). For instance, the transmitted power of mode C (Fig. 2d$_3$) reaches a minimum after the tuning of $S_{11}$ and $S_{12}$, yet the minimization of the overall crosstalk from all the concurrent modes results in an increase of transmission of mode C after the tuning of the last stage $S_{13}$. This means that in practice the mesh configuration cannot necessarily be reliably achieved just using the information provided by external detectors coupled at the output ports, since convergence issues due to local minima can arise, at least if the mesh is not quite perfect. Thus, though such an algorithm based only on overall output powers may work (see the progressive algorithms using only output detectors in [19] Appendix B and in [22], and the global algorithms in [14]), approaches with embedded detectors may offer faster and more robust convergence, in addition to the ability to configure the mesh with all input modes present simultaneously.

Any input mode can be reconstructed at any output port with similar performance. For instance, Fig. 2e shows that, by properly setting $\mathbf{M}_1$, mode reconstruction at port $Out_1$ for any particular chosen input is achieved with less than -20 dB residual crosstalk of the concurrent modes over a wavelength range of about 10 nm. More generally, the mesh transmission matrix

$$\mathbf{H}_{mesh} = \prod_{m=1}^{N-1} \mathbf{M}_{N-m} \tag{4}$$

can be configured to give any desired permutation of inputs to outputs, as in a switching matrix. In other words, the overall matrix of the system $|\mathbf{H}_{mesh}\mathbf{H}|^2$, describing the power transmission of the input modes {A, B, C, D} to the output ports of the mesh, can be chosen to take the form of a generic permutation matrix. This means that not only can the mesh perform an inversion of the **H** matrix, but the reconstructed modes can be also sorted out or switched arbitrarily at the output ports. The results in Fig. 3 show the measured light power at the output ports {$Out_1$ $Out_2$, $Out_3$, $Out_4$} for several configurations of the full mesh. In all the cases considered, the power of the concurrent channels is more than 20 dB below the power of the reconstructed mode. The automatic configuration of the entire mesh is typically achieved in less than 15 seconds (see Methods) and any permutation is allowed.

We note, incidentally, that this performance is achieved even though the intensity split ratio of the directional couplers of the MZIs is quite far from the ideal 50:50 condition (we estimate about 72% coupling in the fabricated device at 1525 nm wavelength), showing the robustness of this approach (see Supplementary Note 5 and Supplementary Fig. 4 for additional discussion). Recent approaches may allow yet further performance optimization even with such imperfect directional couplers[22,33] and/or with broadband couplers[34].

**On-chip MDM channel unscrambling**

To demonstrate the recovery of the information encoded in the optical modes undergoing the mixing process, we injected four data channels (see Fig. 4a), all at 1525 nm wavelength, on the separate fibres to form the inputs A, B, C and D (see Methods and Supplementary Fig. 5 for details on the experimental setup). When the mesh is not configured, mode mixing results in deep time-variations in the spectrum of the optical signal measured at the device output ($OUT_1$ in Fig 4b), because of the coherent beating of the four spectrally overlapped channels. In contrast, the spectrum of the reconstructed channel exhibits only a tiny frequency-domain ripple due to the residual -20 dB crosstalk of the three concurrent channels. The mode mixing leads to complete closure of the signal eye diagram (insets of Fig. 4b), which is effectively restored after mode unscrambling. Panels in Fig. 4c show the eye diagrams of each reconstructed channel at port $Out_1$, with no deterioration as more concurrent channels are switched on. Bit error rate (BER) measurements (Fig. 4d) show a power penalty < 2 dB at a BER of $10^{-9}$ as additional channels are turned on.

The MZI mesh can also self-configure automatically to track modes that are mixed by a time-varying scrambling process. The integrated mode mixer was deliberately perturbed by shining a 980nm-wavelength light beam with an intensity of 3.3 MW/cm$^2$ onto it (see Fig. 5a). Absorption of this light in the silicon of the mode mixer generates free carriers with a density of about $10^{18}$ cm$^{-3}$ (see Methods), and leading to local changes in the refractive index, both from free carrier dispersion (blue shift) and thermal (red shift) effects[35]. These changes affect the self-imaging process along the multimode silicon waveguide[36] and modify the mode mixer behaviour (Fig. 5b). With the 980 nm beam off (Fig. 5b$_1$), the mesh is configured to a reference state where a given channel (A) is reconstructed at one output port ($Out_1$). With the 980 nm beam on, the mode mixer (Fig. 5b$_2$) is sufficiently perturbed to completely impair mode reconstruction if the mesh is not adaptively configured. Figure 5b$_3$ shows that the eye diagram of the output channel is successfully recovered after automatic reconfiguration of the mesh.

## Discussion

We have demonstrated all-optical mode reconstruction, unscrambling and sorting in a silicon photonic circuit using a self-reconfiguring interferometric mesh. Since each mesh element (tuneable beam splitter) is locally monitored and feedback-controlled by transparent detectors, the progressive self-configuration of the mesh is reduced to a repeated two-degrees-of-freedom problem independently of the port-count of the mesh. This feature enables the implementation of simple, accurate and robust control of arbitrarily large meshes for the manipulation of a large number of modes[20]. This overall concept of transparent on-chip monitoring and adaptive feedback control of elementary photonic elements can be extended to arbitrary mesh topologies, such as the ones that have been recently proposed to implement programmable photonic processors[18,37,39-41].

Mode unscrambling on a photonic chip can be also exploited to improve the performance of recently proposed silicon photonics devices for the manipulation of MDM optical channels[6-11], where a one-by-one mapping of the modes of single-mode waveguides to predetermined modes of multimode waveguides is performed, assuming that no mode coupling occurs in the multimode waveguide. In reality, mode mixing could be induced by sharp bending, waveguide crossing, as well as fabrication imperfections, these effects being difficult to predict at design time, and potentially also varying in time in uncooled photonic chips because of the different temperature sensitivities of the different guided modes. The approach presented here can be extended also to other semiconductor photonics platforms, such as InP, where modes are not easily separable because of the similarity of their phase velocities[12] and where the CLIPP operation has also been successfully demonstrated[38]. We expect applications of this approach in mode (de)multiplexers[6-8], multimode switches[9], mode converters[10], switchable mode exchangers[11] and other programmable photonic processors[18,37,39,40] for applications in a variety of different fields, such as telecommunications, imaging, sensing, secrecy, and quantum information processing[5].

**Methods**

**Design and technology of the silicon photonic circuit.** The silicon photonic circuit was realized on a 220-nm SOI platform through a LETI-ePIXfab multi-project-wafer run. Tuneable beam splitters are realized using thermally actuated balanced MZIs, with 40-μm-long directional couplers (300 nm waveguide spacing in the coupling region) and 120-μm-long arms spaced by 20 μm. Titanium nitride integrated heaters, with a width of 1 μm and a length of 100 μm, provide π-phase-shift with an electrical power consumption of about 10 mW. The CLIPPs detectors are made of two 20 μm × 50 μm electrodes, mutually spaced by 100 μm, which are fabricated using the same metal layer as for the heaters. Gold metal strips connect the thermal actuators and the CLIPPs to the 100 μm × 100 μm contact pads, where wire-bonding of the photonic chip to the external electronic circuit is performed. The four modes are coupled to the photonic chip through two arrays of input and output grating couplers, which are spaced by 127 μm. To avoid on-chip differential losses as well as to balance all the interferometric paths of the mesh, folded waveguide sections are added between different stages of the mesh. All the bends throughout the circuits have a curvature radius of 20 μm, allowing very low reflections and negligible bending losses. The circuit footprint, including metal routing and contact pads, is 3.7 mm × 1.4 mm.

**Electronic platform for tuning and locking.** The silicon photonic chip is mounted onto a PCB and is wire-bonded to a specially designed electronic ASIC[29] realized in a 0.35-μm AMS CMOS process. The ASIC contains a low-noise front-end amplifier followed by a fully integrated lock-in system to extract the in-phase and quadrature components of the waveguide impedance[24]. The ASIC has 4 parallel read-out channels, each featuring an 8x input multiplexer to address up to a total of 32 CLIPPs. The 4 output signals from the ASIC are acquired and conditioned by the FPGA-based electronic platform, are digitally demodulated at the frequencies $f_q$ and processed by tuneable IIR filters (down to 4 Hz bandwidth) to identify the power level of each mode. The FPGA drives the 12 heaters of the silicon photonic chip to tune and lock the 6 MZIs to the desired working points. In the experiments, the system has been set to perform the CLIPP read-out in 50 ms, allowing an automatic 2D scan of each MZI (30×30 pixel map, as in Figure 2c) in about 45 seconds. The automatic tuning and locking of the entire mesh is achieved in less than 15 seconds. The reconfiguration time could be considerably reduced with further optimized tuning strategies.

**Experimental setup for on-chip unscrambling of MDM channels.** A detailed schematic of the experimental setup is shown in Supplementary Fig. 5. The four channels encoded on modes {A, B, C, D} are generated by using a common laser source, with emission wavelength of 1525 nm, that is intensity modulated at a data rate of 10 Gbit/s (on-off keying, $2^{31} - 1$ pseudo random bit sequence) through a commercial LiNbO$_3$ Mach Zehnder modulator. After being amplified through an erbium doped fibre amplifier (EDFA), the modulated signal is divided by a 1×4 fibre optic splitter. The four data streams are de-correlated using coils of standard single-mode fibers of different lengths, introducing relative delays (>10 μs) much greater than the signal coherence length. Variable optical attenuators (VOAs) are employed to equalize the channel optical power to 0 dBm at the input of the silicon photonic circuit. Polarization controllers enable the selection of the transverse electric (TE) polarization at the output of the glass transposer (see Fig. 1d) in order to optimize the coupling efficiency of each channel with the optical waveguides. At the output ports of the circuit, the transmitted signals are amplified by an EDFA followed by a filter (0.3 nm bandwidth) that is added to reduce off-band amplified spontaneous emission noise. A VOA is used to control the received power at the input of the photodetector to perform the BER and eye diagram measurements.

**Light-induced perturbation of the mode mixer**. The light beam used to modify the response of the mode mixer is generated by a fibre-coupled laser source, which emits light at 980 nm with a maximum power $P = 20$ dBm. A lensed fibre with a spot area $A$ of about 3 $\mu m^2$ at the focal distance is used to irradiate the mode mixer from the top with an intensity of about 3.3 MW/cm$^2$. Since the spot size $A$ is much smaller than the mode-mixer size (10 $\mu m \times 180$ $\mu m$, see Supplementary Note 2), only a small portion of the device is directly exposed to the light beam. To estimate the density of free carriers $N$ generated in the device, we assume uniform absorption of the light across the core layer. Neglecting the reflection at the air/silica/silicon interfaces, the carrier density is governed by the following rate equation

$$\frac{dN}{dt} = \frac{\alpha P}{A h \nu} - \frac{N}{\tau}, \qquad (5)$$

where $\alpha$ is the silicon absorption coefficient at 980 nm ($\alpha \sim 100$ cm$^{-1}$) and $h\nu$ is the photon energy (1.26 eV). In silicon waveguides, the free carrier lifetime $\tau$ typically ranges from fraction of nanosecond to several tens of nanoseconds[35]. Assuming $\tau = 1$ ns, the steady-state carrier density $N = \frac{\alpha P \tau}{A h \nu}$ is estimated to be of the order of $2 \times 10^{18}$ cm$^{-3}$.

**Acknowledgments.** The research leading to these results received funding from the European Union's Seventh Framework Programme (FP7/2007/2013) under grant agreement n. 323734 BBOI (Breaking the Barriers on Optical Integration) and by Multidisciplinary University Research Initiative grant (Air Force Office of Scientific Research, FA9550-12-1-0024). This work was (partially) performed at Polifab, the micro- and nanofabrication facility of Politecnico di Milano ([www.polifab.polimi.it](www.polifab.polimi.it)). The Authors acknowledge F. Zanetto for the support in measurements.

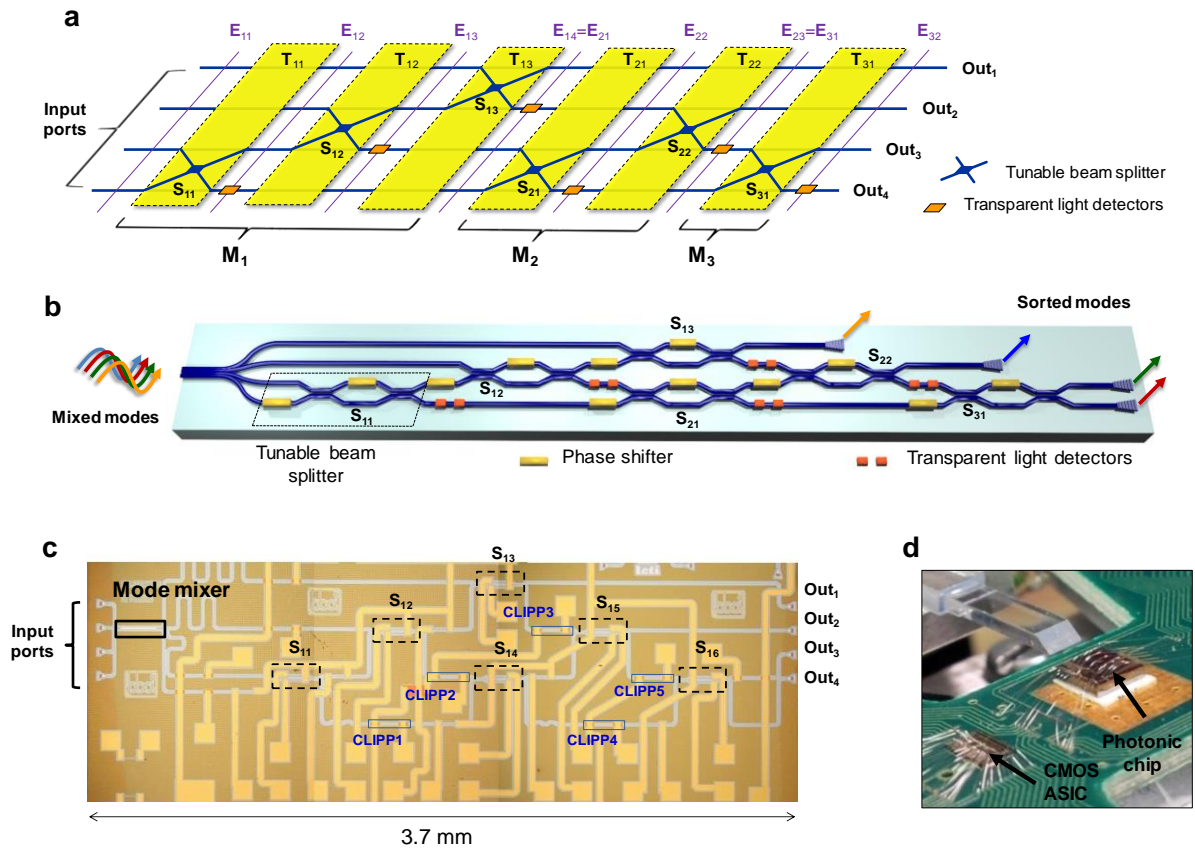

**Figure 1| Self-configuring mode unscrambler integrated in a silicon photonic chip.** (**a**) Schematic concept of an $N \times N$ ($N$=4) triangular mesh of tuneable beam splitters implementing any arbitrary transformation on $N$-dimensional input vectors. Transparent detectors at the output port of each beam splitter monitor the evolution of the optical field $\boldsymbol{E}_{m,k}$ along the entire mesh enabling local control operation on each beam splitter individually. (**b**) Guided-wave implementation of the mesh through a lattice of two-port cascaded MZIs realizing the tuneable beam splitters controlled through a pair of integrated phase shifters. (**c**) Silicon photonic 4-mode unscrambler consisting of six thermally-actuated MZIs individually monitored by transparent CLIPP detectors. Mode scrambling is induced on chip through a multimode waveguide section (*mode mixer*). Self-configuration and stabilization of the circuit is performed through a CMOS ASIC (**d**) bridged to the silicon chip, which is connected to an FPGA controller.

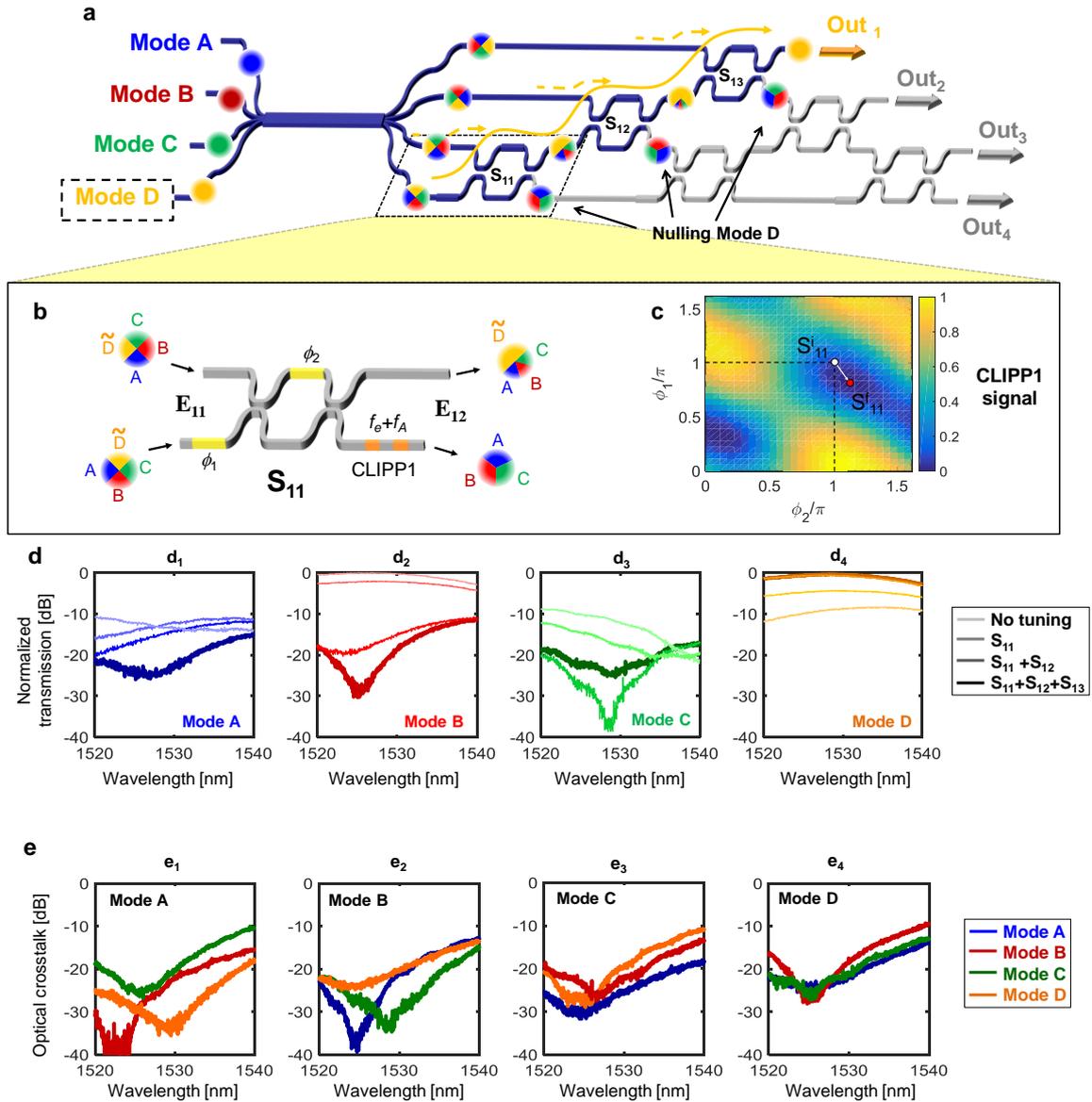

**Figure 2| On-chip unscrambling of optical modes.** Mixed modes are reconstructed at the output port of the 4×4 silicon photonic mesh by sequentially tuning the MZI beam splitters. (**a**) To reconstruct the first mode ($m = 1$, Mode D) at port Out$_1$, the first row of the mesh ($M_1$) is configured by progressively nulling the light intensity at the lower output arms of MZI $S_{11}$, $S_{12}$ and $S_{13}$, where a CLIPP detector is integrated (**b**). (**c**) Normalized power of Mode D measured by CLIPP1 integrated after $S_{11}$. Depending on the initial MZI biasing ($S^i_{11}$), convergence to different equivalent solutions $S^f_{11}$ (local minima of the map) may be achieved. (**d**) Mesh configuration makes the transmission of the mode reconstructed at port Out$_1$ progressively increase (**d$_4$**), while the crosstalk due to the concurrent modes A (**d$_1$**), B (**d$_2$**) and C (**d$_3$**) reduces. (**e**) Reconstruction of mode A (**e$_1$**), mode B (**e$_2$**), mode C (**e$_2$**), and mode D (**e$_2$**) at port Out$_1$ can be achieved with less than -20 dB residual crosstalk of the three concurrent modes over a bandwidth of about 10 nm.

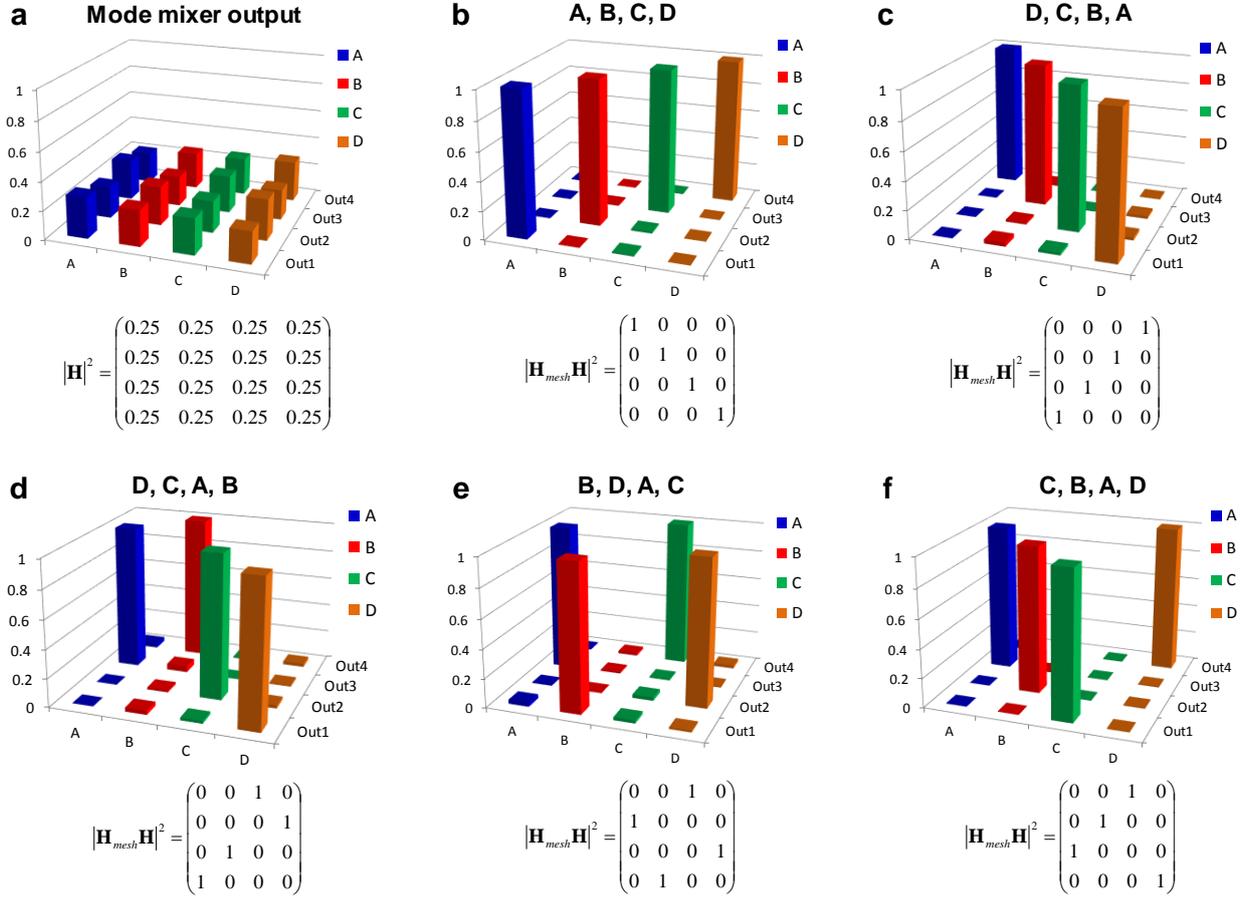

**Figure 3| On-chip mode sorting.** The mesh transmission matrix **H**mesh can be configured in order to sort the reconstructed modes {A,B,C,D} arbitrarily at the output ports {Out$_1$, Out$_2$, Out$_3$, Out$_4$} of the mesh according to any 4x4 permutation power transmission matrix |**H**mesh**H**|$^2$. Given the mode scrambling introduced by the mode mixer **H**, spreading the power of the input modes almost equally in the input waveguides of the mesh (**a**), panels (**b**)-(**f**) show the normalized light power at the output ports of the mesh, when it is configured to extract the modes in the follow order: **b**) A, B, C, D; (**c**) D, C, B, A; (**d**) D, C, B, A; (**e**) C, A, D, B; (**f**) C, B, A, D.

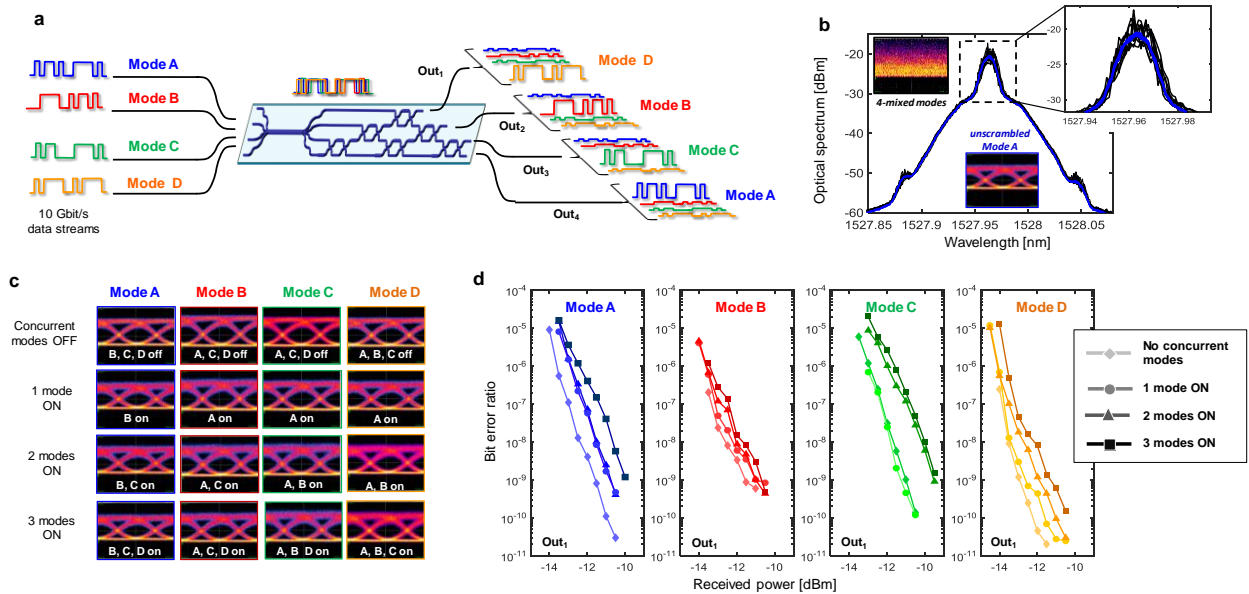

**Figure 4| On-chip unscrambling of MDM optical channels.** Information encoded in four scrambled 10 Gbit/s intensity modulated MDM channels is recovered after mode reconstruction performed by the silicon photonic mesh. (**a**) Schematic of the experimental setup employed for the generation and detection of the four MDM channels (more details in Methods and Supplementary Fig. 5). (**b**) As a consequence of mode mixing, the spectrum of the four mixed channels (black curves) exhibits deep time-varying oscillations, which disappear after mode reconstruction (mode A, blue curves). Displayed curves refer to ten successive measurements taken at output port $Out_1$. The corresponding time domain signals are shown in the eye-diagrams in the insets. Eye diagram (**c**) and BER (**d**) measurements (port $Out_1$) demonstrate that information encoded in each channel can be retrieved with a very small power penalty independent of the number of mixed modes.

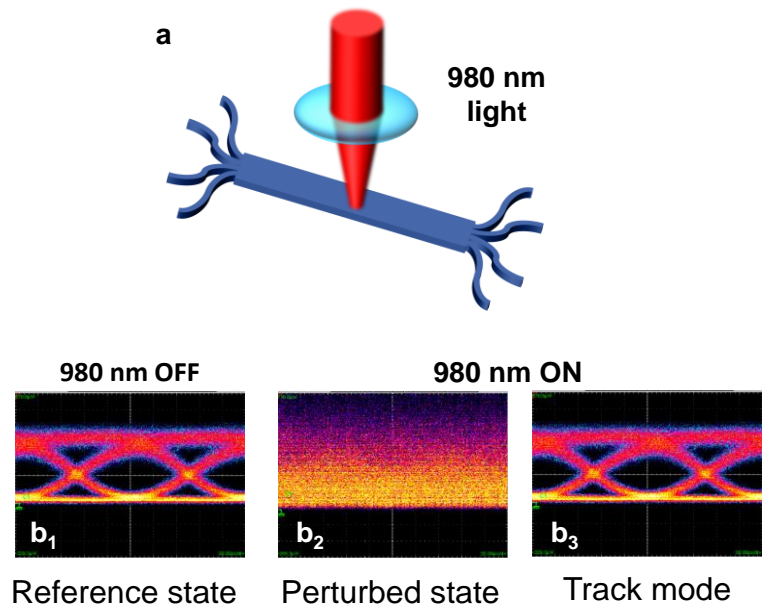

**Figure 5| Reconstruction of modes scrambled by time-varying mixing.** (**a**) A light source (980 nm) is used to perturb the mode mixer integrated in the silicon chip in order to modify the relative amplitude and phase of the mixed modes. (**b**) After configuring the mesh to reconstruct channel A at port $Out_1$ (reference state, **b₁**), the 980 nm source is switched on to modify the mode mixing, thus impairing mode reconstruction at the mesh output (perturbed state, **b₂**). In the track mode (**b₃**), the mesh adaptively self-configures by controlling each MZI through a local feedback loop, in order to automatically compensate against time-varying mixing of the modes.

# Supplementary Information

**Supplementary note 1: Photonic mesh architectures implementing arbitrary linear operations**

At present, three architectures, made up from meshes of 2x2 interferometers, are known that can implement arbitrary unitary transforms between a vector of optical input amplitudes and a corresponding vector of output amplitudes for coherent light at a given wavelength: a "triangular mesh" architecture[15,16, 19,20.23] (which is used in our work here), a "cascaded binary tree" architecture[19], and a "rectangular mesh" architecture[37]. Of these, both "triangular mesh" and "cascaded binary tree" architectures can be configured automatically using "training" vectors of inputs and simple progressive algorithms based on detection and simple one- or two-parameter feedback minimization processes[19,20]. In these "trainable" architectures, a given linear transform is trained using vectors that are the Hermitian adjoints of the desired rows of the corresponding matrix (as we do in this work). All three of these architectures require only a number of phase shifters that corresponds to the number of real numbers required to specify an arbitrary NxN unitary matrix, and so are optimally efficient in that sense.

For non-unitary transforms (i.e., arbitrary matrices), two architectures are known: an architecture based on the singular value decomposition (SVD) of the desired matrix[20], and one based on the use of a 2Nx2N unitary matrix to implement an NxN non-unitary transform by operator dilation[16]. The SVD approach can be "trainable" and has the minimum number of required phase shifters. The SVD approach can be implemented using two unitary transforms and an additional row of modulators[20]. Each such unitary transform can be implemented using any of the above unitary architectures. If "trainable" unitary transform meshes are used, the overall non-unitary function can be trained using appropriate vectors at the inputs for one of the unitary transforms, and by shining appropriate vectors back into the output for training the other unitary transform. Hence the self-configuring approach of our work could also be applied to implement "trainable" non-unitary transformations that mathematically could also undo scattering with different loss on different modes.

**Supplementary note 2: Electronic read-out of the CLIPP monitor**

As reported in previous contributions [11], the CLIPP monitors the light intensity in the waveguide by measuring the light-dependent variation of the conductance $\Delta G$ of the waveguide. Non-invasive monitoring is achieved by remotely performing an impedance measurement without electrically contacting the CLIPP electrodes with the Si core. A top-view picture of one of the CLIPPs integrated in the MZI mesh used in this work is shown in Supplementary Fig. 1a. To by-pass the access capacitance provided by the insulating $SiO_2$ top cladding, the CLIPP electrodes are AC-coupled to the Si waveguide core. A sinusoidal voltage $V_e$ at frequency $f_e$ is applied to one of the CLIPP electrodes, while the current $i_e$ at the other electrode is collected with a synchronous electrical detection architecture. The CLIPP readout requires a low-noise transimpedance amplifier (TIA) and a lock-in detection scheme that are both integrated into the CMOS ASIC connected to the silicon chip. Details on the design of the ASIC can be found in [18].

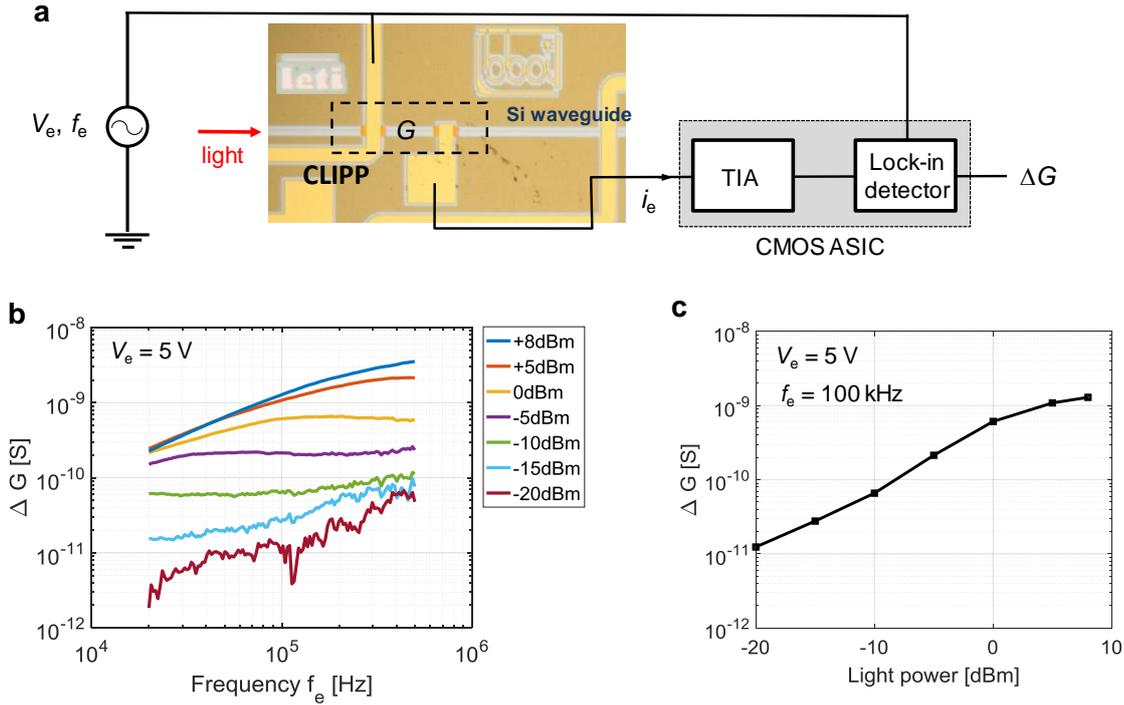

**Supplementary Fig. 1**. Performance assessment of the CLIPP monitor. (a) Top view photograph of the one of the CLIPPs integrated in the MZI mesh and block diagram of the electronic circuit integrated in the CMOS ASIC for the read–out of the CLIPP; (b) Electric signal provided by the CLIPP versus the frequency of the applied voltage signal for increasing optical power in the silicon waveguide. (c) Responsivity curve of the CLIPP measured at a frequency $f_e$ = 100 kHz, where the sensitivity to light variation is maximum.

Supplementary Fig. 1b shows the electrical signal (conductance variation ΔG) provided by a stand-alone test CLIPP fabricated on the same chip of the MZI mesh as a function of the readout frequency $f_e$ for increasing optical power level. Maximum sensitivity to optical power variation is observed around 100 kHz. At this frequency, the responsivity curve of the CLIPP (Supplementary Fig. 2c) shows a sensitivity of at least -20 dBm with a dynamic range of 30 dB. This sensitivity enables accurate monitoring of each MZI tunable beam splitter to achieve mode reconstruction with a -20 dB residual crosstalk.

**Supplementary note 3: Integrated mode mixer**

The integrated mode mixer responsible for mode scrambling consists of a multi-mode waveguide section with four input ($I_1$,…, $I_4$) and four output ($O_1$,…, $O_4$) single mode waveguides, resulting in the multimode interference coupler shown in the schematic of Supplementary Fig. 2a. Electromagnetic simulations based on the Eigenvalue Mode Expansion (EME) method were performed to optimize the design of the mode mixer in order to reduce the loss created by the imperfect self-imaging of the field at the output port of the multimode region (see Supplementary Fig. 2b), since the occurrence of loss in the mode mixer would impair mode orthogonality, thus affecting the mode reconstruction performed by the MZI mesh. To reduce the loss, the single-mode input/output waveguides are linearly tapered up to a width of 2 μm. In the circuit presented in this work, the mode mixer integrated before the MZI mesh is 180 μm long and 10 μm wide. Supplementary Fig. 2c shows the spectral response of a stand-alone mode mixer that was fabricated on the same chip for testing purposes. When the light is injected from one input port (due to the symmetry of the device, only curves referring to inputs $I_1$ and $I_2$ are shown) an almost-wavelength-independent 25% (± 2%) mode splitting is observed at all four output ports, thus maximizing the mode scrambling between the input modes. The overall insertion loss of the mode mixer was evaluated by comparing the sum of the power leaving the output ports to the power collected from a reference straight waveguide; for every input port an excess insertion loss lower than 0.7 dB was estimated, thus confirming that mode scrambling is performed without impairing mode orthogonality.

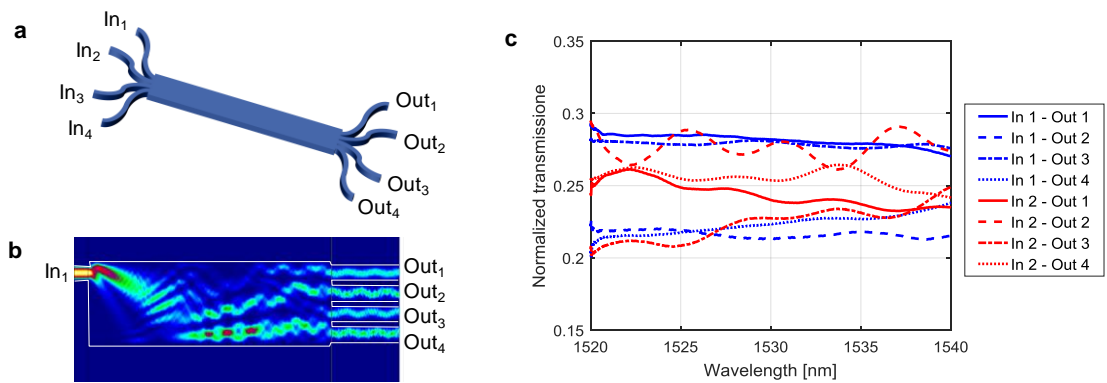

**Supplementary Fig. 2**. Optical characterization of the integrated mode mixer. (a) Schematic and (b) electromagnetic simulation of the mode mixer. (c) The fabricated mode mixer splits the input power of each input mode to all four output ports with a 25% (± 2%) split ratio over the 1520 - 1540 nm wavelength range considered in this work.

**Supplementary note 4: Mode labelling and identification with modulation tones**

When mixed modes are simultaneously injected into the MZI mesh, the CLIPP detectors can identify the power associated with each mode regardless of the presence of other concurrent modes injected at other input ports and scrambled by the mode mixer. To enable mode discrimination, each mode is labelled with a weak "key" or pilot tone before being coupled to the silicon chip. In previous reports we have demonstrated that such a labelling operation can be performed without affecting the quality of the signals [19, 20].

Pilot tones with a 5% peak-to-peak relative intensity are generated through external MZI lithium niobate modulators which are biased at the linear working point (3 dB attenuation). The tone frequency $f_q$ = {4 kHz, 7 kHz, 10 kHz, 11 kHz} of the $q$-th mode ($q$ = A, B, C, D) was suitably chosen to avoid mutual overlap of the overtones that can be generated by non-perfectly-linear response of the modulators. To identify the $q$-th mode, the CLIPPs are demodulated twice, first at the read-out frequency $f_e$ around which the CLIPP sensitivity to optical power variations is maximized (about 100 kHz in the reported experiments, see Supplementary Note 1), then at the frequency $f_q$, of the mode to be monitored. Second demodulation at a frequency different from $f_q$ produces a very low crosstalk signal (lower than −50 dB), which is mainly due to the noise level of the electronic front-end [20].

The effectiveness of the mode identification performed by the CLIPP and its use for the monitoring of the tunable beam splitters of the mesh is shown in Supplementary Figure 3. The three maps show the signal provided by CLIPP1 when the beam splitter $S_1$ is tuned by changing the phases $\phi_1$ and $\phi_2$. With respect to the case where only one mode (D) is injected in the mesh (a), the presence of concurrent channels strongly modifies the map [in (b) also channel B is switched on], hindering the biasing of the MZI at the proper working point for mode D reconstruction. Mode labelling through pilot tones (c) enables monitoring and control of the state of the MZI with no side effects associated with the presence of the concurrent channels.

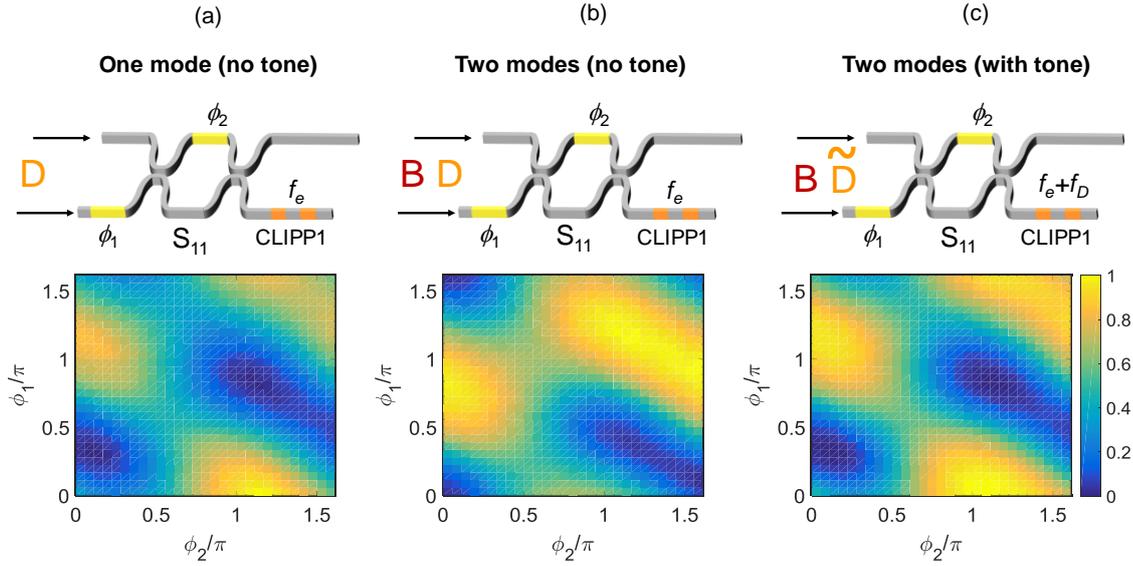

**Supplementary Fig. 3**. CLIPP-assisted monitoring of the tuneable beam splitters of the mesh by using mode labelling. Maps show the signal measured by CLIPP1 during the tuning operation of the beam splitter $S_1$ as a function of $\phi_1$ and $\phi_2$, when: (a) only mode D is injected in the mesh (concurrent modes off, no tone applied); (b) concurrent mode B is switched on (no tone applied, CLIPP frequency $f_e$); (c) concurrent mode B is switched on ($f_D$ tone applied on mode D, CLIPP frequency $f_e + f_D$).

**Supplementary note 5: Tolerance analysis of mode reconstruction**

Numerical simulations were performed by using the transmission matrix method (TMM) to investigate the sensitivity of the mesh to fabrication imperfection in the directional couplers of the MZIs. Supplementary Figure 4 shows the overall crosstalk, averaged over a bandwidth of 10 nm around 1525 nm, that is provided by the other three concurrent channels when channel A (blue), B (red), C (green), and D (yellow) are respectively reconstructed at the output port $Out_1$. Results are reported only for split ratios > 0.5 because crosstalk curves are symmetrical with respect to the ideal condition (3 dB directional coupler). A crosstalk lower than -25 dB is observed up to a split ratio of about 0.75, thus implying that no significant crosstalk degradation occurs for relative deviations as large as 50% from the ideal condition.

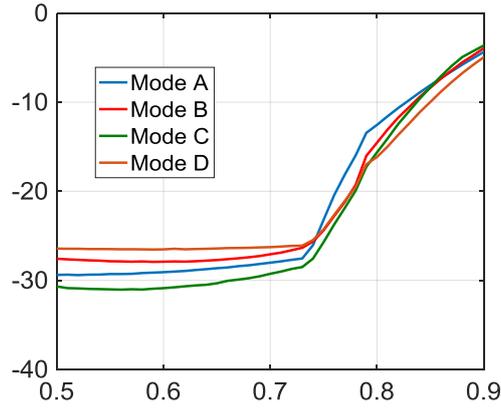

**Supplementary Fig. 4**. Robustness of mode reconstruction versus fabrication tolerances in the directional coupler of the mesh. Curves show the simulated crosstalk given by the all the concurrent channels when mode A (blue), B (red), C (green), and D (yellow) is reconstructed at the port $Out_1$. No significant crosstalk degradation is observed up to a split ratio of about 0.75.

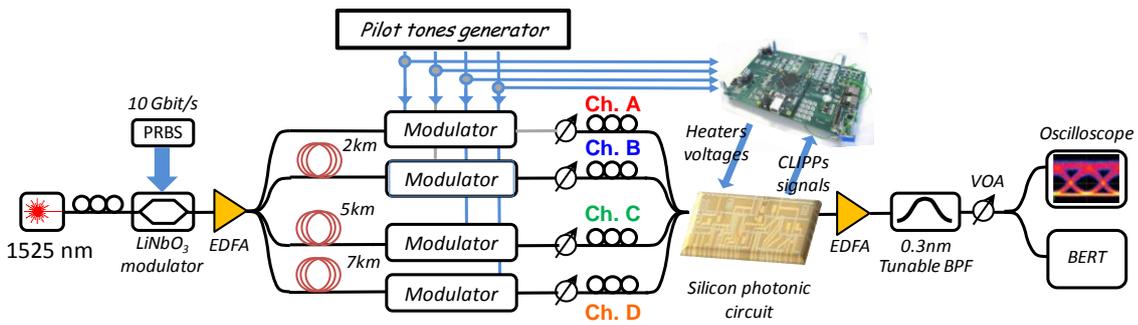

**Supplementary Fig. 5**. Experimental setup employed for the demonstration of all-optical unscrambling of mixed MDM channels.